\pdfoutput=1

\documentclass[11pt]{article}
\usepackage{jheppub}

\usepackage{amsmath}
\usepackage{amssymb}
\usepackage{graphicx,color,slashed,braket}
\usepackage{bbm}
\usepackage{ifpdf}
\usepackage{comment}
\usepackage{soul}
\usepackage{parskip}
\usepackage{booktabs}
\usepackage{multirow}
\usepackage{makecell}
\usepackage{float}
\usepackage{verbatim}
\usepackage{caption}
\usepackage{cancel}
\usepackage{enumerate}
\usepackage{tcolorbox}
\usepackage[font=small,labelfont=bf]{caption}

\usepackage{color}
\definecolor{dark-gray}{gray}{0.20}
\definecolor{gray}{gray}{0.30}
\definecolor{light-gray}{gray}{0.80}
\definecolor{dark-red}{rgb}{0.7,0,0}
\definecolor{dark-green}{rgb}{0.1,0.4,0}
\definecolor{dark-blue}{rgb}{0.3,0.3,0.7}
\definecolor{light-blue}{rgb}{0.8,0.8,1}

\usepackage{tikz}
\usetikzlibrary{calc}

\usepackage{hyperref}
\usepackage{setspace}
\hypersetup{
	colorlinks=true,
	linkcolor=dark-blue,
	citecolor=dark-red,
	urlcolor=dark-green,
	linktoc=page
}
\newcommand{\be}{\begin{equation}}
\newcommand{\ee}{\end{equation}}

\def\be{\begin{equation}}
\def\ee{\end{equation}}
\def\bea{\begin{eqnarray}}
\def\eea{\end{eqnarray}}

\newcommand{\Ra}{R_{\rm{AdS}}}

\allowdisplaybreaks

\title{A note on the holographic consistency of DGKT-type vacua with $h^{2,1}=0$}

\author[1,2]{Filippo Revello}
\author[1]{and Farah Verbeure}

\affiliation[1]{Instituut voor Theoretische Fysica, KU Leuven, Celestijnenlaan 200D, B-3001 Leuven, Belgium}
\affiliation[2]{Leuven Gravity Institute, KU Leuven, Celestijnenlaan 200D, B-3001 Leuven, Belgium}

\emailAdd{filippo.revello@kuleuven.be}
\emailAdd{farah.verbeure@student.kuleuven.be}

\notoc

\abstract{Recent works have pointed out the existence of a holographic constraint on (extremal) three-point functions of scalar moduli in scale-separated AdS vacua. Moreover, it has been shown that this constraint is satisfied in the DGKT scenario in massive type IIA for the original $\mathbb{T}^6/(\mathbb{Z}_3 \times \mathbb{Z}_3)$ orbifold, as a result of a series of unexpected cancellations. We extend the analysis to more elaborate scenarios, involving geometries with less symmetry and more complicated triple-intersection numbers. Surprisingly, the cancellations persist in all examples with
$h^{2,1}=0$, revealing a universal structure in the cubic
couplings and the corresponding mass eigenstates.
These results subsequently motivated and provided essential
ingredients for a general proof of the cancellation for
arbitrary Calabi--Yau orientifolds with $h^{2,1}=0$.}

\begin{document}

\maketitle

\newpage
\tableofcontents
\vspace{0.5cm}
\hrule
\section{Introduction}
One of the most fundamental requirements for a string compactification to describe our observed universe is a large hierarchy between the size of the internal dimensions\footnote{A more precise definition is given by the lowest (non-zero) eigenvalue of the Laplacian on the compact manifold, corresponding to the mass of the lightest Kaluza-Klein (KK) mode.} and the characteristic scale of the non-compact spacetime (\emph{i.e.} the Hubble radius). Such a property is known as \emph{scale-separation}, and it is a necessary condition for the existence of a consistent Effective Field Theory (EFT) description which is inherently lower-dimensional. However, the existence of controlled, scale-separated solutions in String Theory/Supergravity is a particularly debated topic, see \cite{Coudarchet:2023mfs} for a review. In the case of Anti-de Sitter (AdS) vacua it is a particularly well-posed problem, which can be phrased in terms of vacua with a parametric separation between the AdS radius and the KK scale, $R_{\rm{AdS}} \gg R_{KK}$. Regardless of the phenomenological implications, it can also be seen as a fundamental question in the context of holography, \emph{i.e.} whether there exist any examples of the AdS/CFT duality involving a geometry with a compact spacetime factor whose radii are much smaller than the typical AdS scale. Moreover, the dual holographic picture would be in terms of CFTs with a parametric gap in the spectrum of primary operators, of which no examples are currently known. 

At the classical level, there exist known obstructions to obtaining scale-separation in supergravity constructions using only fluxes and positive energy sources, which can be understood from the difficulty in decoupling the internal and external curvatures \cite{Gautason:2015tig}. Moreover, Swampland arguments suggest that scale-separated solutions may only arise in the presence of minimal or no supersymmetry ~\cite{Cribiori:2020use,Cribiori:2022trc,Montero:2022ghl,Cribiori:2024jwq} (see however \cite{Cribiori:2026caf} in the special case of 2d). Known proposals to achieve scale separation in String Theory can roughly be divided into two classes. The first kind, of which the KKLT \cite{Kachru:2003aw} and LVS \cite{Balasubramanian:2005zx,Conlon:2005ki} scenarios are the most famous examples, rely on the presence of quantum corrections, and will not be discussed in this work. The second class, usually referred to as classical, only employs classical ingredients as well as negative tension sources such as O-planes. The best known examples are given by the DGKT vacua of \cite{DeWolfe:2005uu} (see also \cite{Camara:2005dc}) in massive type IIA with O6 planes. They realise scale separation because certain fluxes are not constrained by tadpole considerations and can therefore be taken to be arbitrarily large, $F_4 \sim \bar{e} \gg 1$. Over the years, similar AdS$_4$ constructions have been obtained in the case of massless type IIA \cite{Cribiori:2021djm,Carrasco:2023hta,VanHemelryck:2024bas}, as well as AdS$_3$ vacua in type IIA \cite{Farakos:2020phe,Farakos:2023nms,Farakos:2025bwf,Tringas:2026ncg}, type IIB \cite{VanHemelryck:2025qok,Miao:2025rgf} and more recently in heterotic string theory \cite{Tringas:2025bwe}. While such solutions claim to achieve scale separation in the controlled limit of large volumes and small string coupling (obtained when $\bar{e} \gg 1$), a particular point of contention arises from the treatment of localised sources in the supergravity equations of motion. Indeed, all of the above constructions rely on the existence of O-planes wrapping cycles along the compact dimensions, whose charge density is \emph{smeared} along the compact dimensions in order to obtain approximate solutions to the classical equations of motion. While this procedure is justified a posteriori in the case of small extra dimensions, it is not completely clear whether the backreaction of the O-planes can be consistently neglected \cite{Baines:2020dmu,Marchesano:2020qvg, Emelin:2022cac}. While recent works have shown this to be the case at the first non-trivial order in a $1/\bar{e}$ expansion \cite{Junghans:2020acz,VanHemelryck:2024bas}, it appears difficult to achieve a definitive consensus using perturbative techniques only. Moreover, possible pathologies related to intersecting O-planes at the orbifold singularities have also been raised (see \cite{Junghans:2023yue} for a resolution in specific examples).

For these reasons, it is tempting to approach the problem from a different angle, using the Anti-de Sitter/Conformal Field Theory (AdS/CFT) correspondence to frame the question in terms of the consistency of the would-be dual CFT. General features of CFT duals to scale-separated vacua have been analysed in \cite{Polchinski:2009ch,El-Showk:2011yvt,Alday:2019qrf,Collins:2022nux}, and possible constraints were suggested in \cite{Bobev:2023dwx,Perlmutter:2024noo}. A similar approach to well-known scenarios of moduli stabilisation (including \emph{e.g.} KKLT and LVS) was also advocated in \cite{Conlon:2018vov,Conlon:2020wmc}. For DGKT specifically, early work along these lines first appeared in \cite{Aharony:2008wz}, establishing a peculiar growth of the central charge with respect to the ``large-N" parameter. Moreover, it was shown in \cite{Conlon:2021cjk,Apers:2022zjx,Apers:2022tfm} (see also \cite{Herraez:2018vae,Marchesano:2019hfb,Apers:2022vfp}) that the low-lying sector of the putative CFT dual to the DGKT construction exhibits a universal spectrum, consisting only of integer scalar dimensions. This was later verified for the 3d constructions of \cite{Arboleya:2024vnp,Farakos:2025bwf, VanHemelryck:2025qok} as well. Finally, it is worth mentioning that a complete realisation of DGKT or its lower-dimensional cousins in terms of the worldvolume theory living on a brane configuration is still lacking, but partial progress has been made in \cite{Montero:2024qtz,Apers:2025pon,Apers:2026lgi,Bedroya:2025ltj,Arboleya:2026irl}.

More recently, it was shown that scale separated vacua describing a genuine EFT in AdS have to obey a holographic consistency condition on the 3-point functions \cite{Bobev:2025yxp}. The latter states that certain cubic couplings in the theory, corresponding to so-called extremal arrangements, must always vanish. In \cite{Bobev:2025yxp}, it was shown how the constraint is verified in the simplest example of a DGKT vacuum, based on the orientifold $\mathbb{T}^6/ \mathbb{Z}_3 \times \mathbb{Z}_3$. On the other hand, Ref. \cite{Revello:2026eqp} found that the constraint can often be violated for compactification geometries with complex structure moduli. The goal of this paper is to explore whether similar cancellations still persist in more complicated orbifolds of DGKT with no complex-structure deformations, \emph{i.e.} $h^{2,1}=0$, and to identify the structures underlying these cancellations. As we will see, the explicit examples reveal a remarkably universal pattern, not only in the vanishing of the effective cubic couplings, but also in the individual contributions from the scalar potential and derivative interactions. In particular, the explicit diagonalisation procedure revealed the structure of the mass eigenvectors, which subsequently played a crucial role in establishing the general result for arbitrary Calabi--Yau orientifolds with $h^{2,1}=0$~\cite{Revello:2026yla}.

The article is structured as follows. In section \ref{sc:dgkt}, we review the known properties of the putative holographic dual to DGKT, as well as the cubic couplings constraint from Ref.~\cite{Bobev:2025yxp}. Explicit models are analysed in section \ref{sc:mods}, as well as the appendices. In particular, we will cover all DGKT orientifolds with $h^{2,1}=0$ appearing in the classification of \cite{Flauger:2008ad}. Finally, we conclude with a brief discussion in section \ref{sc:conc}.

\section{Holographic constraints for DGKT}\label{sc:dgkt}
\subsection{DGKT and its putative holographic dual}\label{ssc:dgkt}
The DGKT vacua \cite{DeWolfe:2005uu} are perhaps the most studied examples of AdS scale-separated vacua built only from classical ingredients, \emph{i.e.} fluxes and O-planes. They arise from massive IIA string theory/supergravity compactified on Calabi-Yau orientifolds with O6 planes. While fully explicit examples arise from simple toroidal orbifolds (such as the original $\mathbb{T}^6/\mathbb{Z}_3\times \mathbb{Z}_3$), the construction can be formulated for a general Calabi-Yau orientifold $X$ admitting a consistent involution, and subject to certain other conditions (\emph{e.g.}~allowing for tadpole cancellation). 

At the level of the 4-dimensional EFT, they can be conveniently described in terms of $\mathcal{N}=1$ supergravity, \cite{Grimm:2004uq,Grimm:2004ua}. For reasons that will be described later, we will only consider models with no complex-structure moduli, and already from here we specialise to $h^{2,1}=0$. The K\"ahler form $J$ and $B_2$ can be expanded in terms of the odd cohomology $w_a \in H^{1,1}_-(X)$, as
\begin{equation}
    J = v^a(x) w_a \qquad B_2 = b^a(x) w_a \qquad \text{where} \qquad  a = 1, \dots, h^{(1,1)}_-,
\end{equation}
which combine into the $\mathcal{N}=1$ chiral multiplets as $t^a= b_a +i v_a$. The K\"ahler potential for the $t^a$'s is given by
\begin{equation}\label{eq:kap}
    K_K = - \text{log}\left(8 \mathcal{V} \right),
\end{equation}
where the expression for the volume is determined by the triple intersection numbers $\kappa_{abc}$ as follows,
\begin{equation}
     \mathcal{V} = \frac16 \kappa_{abc} v^a v^b v^c, \qquad \qquad \kappa_{abc} = \int w_a \wedge w_b \wedge w_c.
\end{equation}
The K\"ahler potential determines the kinetic terms for the moduli as
\begin{equation}\label{eq:kt}
    \mathcal{L}_{\rm{kin}}= -2 K_{I \bar{J}} \partial_{\mu} \Phi^I \partial^{\mu} \bar{\Phi}^{\bar{J}},
\end{equation}
where the factor of 2 is unconventional, and only necessary to make contact with the notation of \cite{Bobev:2025gzu,Bobev:2025yxp}. If $h^{2,1}=0$, the only remaining modulus is given by the axio-dilaton $N=\xi+i z$, with K\"ahler potential
\begin{equation}
    K_Q = -4 \log(2 z) = 4D,
\end{equation}
where $D$ is the 4d dilaton. Finally, one can generate a scalar potential by turning on the fluxes,\footnote{$F_2$ flux can also be turned on, but it only shifts the expectation value of the axions without changing any of the couplings, so it will be omitted.}
\begin{equation}
    F_0 = m_0 \qquad F_4 = e_a \Tilde{w}_a \qquad F_6 =e_0 \qquad H_3=- p \beta_0,
\end{equation}
where $\tilde{w}_a$ is a symplectic basis of $H^{2,2}_-(X)$ and $\beta_0$ represents the only cohomology class of $H^3(X)$. Their effect is encoded in the superpotential
\begin{equation}\label{eq:W}
    W= e_0 + e_a t_a  - \frac{m_0}{6}\kappa_{abc}t_a t_b t_c -2pN,
\end{equation}
polynomial in the chiral superfields. The latter generates a scalar potential given by the usual $\mathcal{N}=1$ formula
\begin{equation}
    V= 2 e^K \left(K^{I \bar{J}} D_{I} W D_{\bar{J}} \bar{W} -3 |W|^2\right),
\end{equation}
where the (unusual) factor of $2$ has again been inserted to make contact with the conventions of \cite{Bobev:2025gzu,Bobev:2025yxp}. This is sufficient to stabilise all of the K\"ahler moduli and dilaton, as well as their associated axions, ensuring parametric separation between the KK and AdS scales.

The putative CFT duals to the (supersymmetric) DGKT vacua enjoy various puzzling properties. As all holographic CFTs, their spectrum contains a parametric gap between a finite number of operators with conformal dimension $\Delta \sim \mathcal{O}(1)$, dual to the moduli, and towers of operators with $\Delta_{KK} \gtrsim R_{\rm{AdS}}/R_{KK} \gg 1$, dual to the KK modes. Moreover, in the case of DGKT, the low-lying part of the spectrum is universal, independently of any details of the compactification geometry \cite{Apers:2022tfm}. For any orientifold with $h^{2,1}=0$, the conformal dimensions of the saxions and axions in the large $\bar{e}$ limit are given by the integer numbers 
\begin{equation}\label{eq:deltadgkt}
    \Delta^s_0= 10, \qquad \Delta_{i}^s=6 \qquad \qquad \Delta^a_0= 11, \qquad \Delta_{i}^a=5 \qquad \qquad i=1,\dots,h^{1,1}_-.
\end{equation}
In particular, such vacua correspond to so-called \emph{dead-end} CFTs, with no marginal or relevant operators. In 3 dimensions, it was shown that no perturbative dead-end CFTs can exist \cite{Nakayama:2015bwa}. Although we will not consider such examples in this note, for $h^{2,1} \neq 0$ the conformal dimensions of the operators dual to the complex-structure moduli are also integer, and given by the numbers
\begin{equation}
    \Delta^s_i=2 \qquad \qquad \qquad \Delta^a_i=3 \qquad \qquad \qquad i=1,\dots,h^{2,1}.
\end{equation}
As can be verified in the formulas above, $3d$ $\mathcal{N}=1$ supersymmetry implies that the conformal dimensions of an axion and a saxion in the same multiplet will satisfy $|\Delta_a-\Delta_s|=1$ \cite{Cordova:2016emh}. The bottom component determines the parity of the associated superconformal multiplet, which is even for a saxion and odd for an axion (see also Appendix \ref{app:3d1}).

\subsection{A cubic coupling constraint}\label{ssc:cctr}

In \cite{Bobev:2025yxp}, it was suggested that families of AdS vacua admitting parametric scale-separation should obey a non-trivial consistency condition, in the form a constraint on the cubic couplings of scalar moduli. Such a constraint applies (but is not limited to) genuine AdS EFTs with a single UV cut-off and a finite number of fields up to spin 2.\footnote{Counterexamples to \eqref{eq:ctr} which do not satisfy all of the assumptions have been presented in \cite{Chester:2025wti,Chester:2025jxg} (see also \cite{Castro:2024cmf}). } The parametric aspect is also crucial, as it allows to identify a large parameter (typically some unbound flux number) that can be mapped to the large central charge $c$ of the (putative) dual holographic CFT. From this point of view, the constraint can be interpreted as a consequence of the large-$c$ factorisation properties of the dual CFT correlators.

More in detail, the constraint applies whenever the dual CFT contains \emph{extremal} arrangements in the spectrum. The latter is defined as a set of three (scalar) primary operators $\left\{\mathcal{O}_i, \mathcal{O}_j,\mathcal{O}_k \right\} $ whose conformal dimensions add up to each other, namely\footnote{Similar constraints also hold for so-called \emph{super-extremal} arrangements with $\Delta_i= \Delta_j + \Delta_k + 2n$, $n \in \mathbb{N}^+$.}
\begin{equation}\label{eq:ext}
    \Delta_i= \Delta_j + \Delta_k.
\end{equation}
As has been known since the advent of AdS/CFT, whenever a holographic CFT contains a triplet of scalar primaries obeying \eqref{eq:ext} \emph{exactly} the corresponding three-point function must vanish \cite{DHoker:1999jke,Aprile:2020uxk}. 

In the situation we are interested in, on the other hand, the $\Delta_i$'s in \eqref{eq:ext} are not protected by supersymmetry and will generically be subject to quantum corrections. The holographic constraint we will present is the generalisation of the previous statement on three-point functions to the case where \eqref{eq:ext} is satisfied at leading order in a $1/c$ expansion.

Let us then consider theories where \eqref{eq:ext} is satisfied by the leading order conformal dimensions $\Delta_i^{(0)}$'s in a large-$c$ expansion. In practice, the $\Delta_i^{(0)}$'s are extracted from the usual AdS/CFT dictionary through the expression
\begin{equation}
    \Delta_i^{(0)} (\Delta_i^{(0)} -d)= m_i^2 R^2_{\rm{AdS}},
\end{equation}
where $m_i$ is the tree-level mass of the dual scalar field $\varphi_i$ in AdS. While such an extremality condition seems to require a certain degree of fine-tuning, we notice that it is often verified in theories with an integer spectrum of conformal dimensions, which is a rather common feature of known scale-separated vacua. 

To illustrate the precise form of the constraint, let us then specialise to the scalar sector of a generic AdS solution, whose scalar excitations can be expanded around the vacuum as\footnote{in Euclidean signature.}
\begin{equation}\label{eq:lag}
\mathcal{L}_{\mathrm{eff}}=  -R-\frac{d(d-1)}{L_{{\rm AdS}}^2}+\frac{1}{2}\left(\partial_\mu \varphi_i\right)^2+\frac{1}{2} \frac{m_i^2}{L_{\rm AdS}^2} \varphi_i^2  +d_{i j k} \varphi_i \partial_\mu \varphi_j \partial^\mu \varphi_k+\frac{c_{i j k}}{L_{\rm AdS}^2} \varphi_i \varphi_j \varphi_k+\ldots
\end{equation}
Let us also define an effective coupling 
\begin{equation}
c_{i j k}^{\prime}=c_{i j k}+\frac{m_i^2-m_j^2-m_k^2}{2} d_{i j k},
\end{equation}
which includes contributions from both derivative interactions and the scalar potential, and is invariant under modifications of the lagrangian \eqref{eq:lag} by some total derivative. On the CFT side, the $c'_{ijk}$ encode the three-point function coefficients $C_{ijk}$ through 
\begin{equation}\label{eq:Crel}
    C_{ijk} = \left(\frac{2}{M_P^2}\right)^2
\frac{\Gamma\left(\Theta_i\right) \Gamma\left(\Theta_j\right) \Gamma\left(\Theta_k\right) \Gamma\left(\Theta-\frac{d}{2}\right)}{2 \pi^d \Gamma\left(\Delta_i-\frac{d}{2}\right) \Gamma\left(\Delta_j-\frac{d}{2}\right) \Gamma\left(\Delta_k-\frac{d}{2}\right)} \, c'_{ijk},
\end{equation}
where $\Theta=\frac{\Delta_i+\Delta_j+\Delta_k}{2}$, $\Theta_i=\Theta-\Delta_i$ and $M_P$ is the Planck mass in $d+1$ dimensions.\footnote{Indeed, the RHS of \eqref{eq:Crel} diverges in the case of a (super)extremal arrangement.}

The central claim of \cite{Bobev:2025yxp} is that for scalar fields $\left\{ \varphi_i,\varphi_j,\varphi_k\right\}$ giving rise to a \emph{(super)-extremal arrangement} at leading order in a $1/c$ expansion, the effective cubic coupling must vanish identically, \emph{i.e.}
\begin{equation}\label{eq:ctr}
    c'_{ijk}=0.
\end{equation}
The rest of this note will be devoted to studying \eqref{eq:ctr} in models of interest, as detailed in the next subsection.

In the specific case of a $3d \, \mathcal{N}=1$ SCFT, dual to $\mathcal{N}=1$ AdS$_4$ in the bulk, an additional simplification arises from supersymmetry. In a supersymmetric theory, three-point functions involving operators belonging to the same superconformal multiplets are not independent, but obey non-trivial relations. This is analysed in Appendix \ref{app:3d1}. In particular, the CFT three-point function coefficients of scalar operators $T_i$ and $B_i$,top and bottom components of the same superconformal multiplet $\Phi_i$, are related by
\begin{equation}\label{eq:susyrel}
    C_{T_i T_j B_k}= f\big(\left\{\Delta_l \right\} \big)\, C_{B_i B_j B_k} \qquad \quad C_{T_i T_j T_k}= g\big(\left\{\Delta_l \right\} \big)\, C_{T_i B_j B_k},
\end{equation}
where $f,g$ are functions of the external conformal dimensions whose expressions are reported in Appendix \ref{app:3d1}. For the multiplets relevant to DGKT, this gives, in particular,
\begin{equation}\label{eq:csusy}
C_{a_0\phi_i a_j}
=
\frac{3}{\sqrt{38}}\,
C_{\phi_0 a_i a_j},
\end{equation}
where the operators have dimensions $(11,6,5)$ and $(10,5,5)$, respectively. Thus, the two extremal arrangements are not independent as the vanishing of the latter three-point function implies the vanishing of the former. More generally, the superspace relations derived in Appendix~\ref{app:3d1} reproduce the ratios between non-extremal three-point coefficients obtained holographically in~\cite{Bobev:2025yxp}, providing a superconformal explanation for these relations. Similar identities have appeared in the context of the $3d$ $\mathcal N=1$ superconformal bootstrap~\cite{Rong:2018okz}.

\subsection{Application to DGKT-type vacua on toroidal orientifolds}

The AdS vacua of the DGKT type in IIA massive supergravity, described in Section \ref{ssc:dgkt}, are a natural setting in which to explore the condition \eqref{eq:ctr}. This is thanks to the natural appearance of a wealth of extremal arrangements.\footnote{We do not need to consider extremal arrangements with an odd number of axions, as they automatically vanish thanks to parity symmetry.} Two particularly important ones are given by the triplets of scalars with dimensions $(10,5,5)$ and $(11,6,5)$, which arise in the K\"ahler sector and are always present. If $h^{2,1}\neq 0$, many more arrangements can arise, including super-extremal ones.

In \cite{Bobev:2025yxp}, the condition \eqref{eq:ctr} was verified for the simplest incarnation of DGKT, obtained by compactifying on an orbifold of $\mathbb{T}^6/\mathbb{Z}_3 \times \mathbb{Z}_3$. Moreover, the conclusion was found to hold both for the usual supersymmetric vacua, as well as the non-supersymmetric ones\footnote{Obtained by changing the sign of some $F_4$ fluxes with respect to the susy vacuum.}. This is a rather surprising result, suggesting the existence of non-trivial structure which is not immediately visible from the bulk point of view. Indeed, the cancellation arises from a careful balancing of the terms in the potential ($c_{ijk}$'s) and the derivative interaction ($d_{ijk}$'s), and holds for any choice of the microscopic parameters (fluxes, intersection numbers, etc). On the other hand, the $\mathbb{T}^6/\mathbb{Z}_3 \times \mathbb{Z}_3$ construction might be regarded as a rather special case, as it enjoys a certain degree of simplicity as well as of symmetry - for example, each $\mathbb{T}^2$ can be mapped to another if the corresponding fluxes are also exchanged. It is then natural to wonder if such an intricate cancellation will again take place for more complicated compactification geometries. In \cite{Revello:2026eqp}, it was shown that models with $h^{2,1} \neq 0$ typically lead to a violation of the constraint, thanks to the presence of many new extremal and super-extremal arrangements in the spectrum. This raises the natural question of whether the constraint might hold universally for compactifications on Calabi--Yau orientifolds with $h^{2,1}=0$. In this work, we address this question through explicit computations in increasingly complicated toroidal examples. Besides providing non-trivial evidence for such a universal statement, these examples allow us to identify structures that are obscured in the general formulation. In particular, we found that not only the final cancellation but also the individual derivative and potential couplings display a universal pattern, despite their complicated dependence on the compactification parameters at intermediate stages.

The universal structure uncovered in these explicit examples subsequently provided the starting point for the general analysis of~\cite{Revello:2026yla}. In particular, the form of the mass  eigenvectors suggested by these calculations proved crucial to establishing the cancellation for arbitrary Calabi--Yau orientifolds with $h^{2,1}=0$.
\begin{table}[h!]
\centering
\begin{tabular}{|c|c|c|c|c|}
\hline  Model &  $h_{+}^{1,1}$ & $h_{-}^{1,1}$ & $\kappa, \hat{\kappa}$ &  Quotients \\
\hline I  & 0 & 3 & $\kappa_{123}=1$ &  $
\begin{array}{c}
\mathbb{Z}_7, \mathbb{Z}_{8-I}, \mathbb{Z}_{12-I}, 
\mathbb{Z}_2 \times \mathbb{Z}_{6^{\prime}}, \mathbb{Z}_3 \times \mathbb{Z}_3,\\
\mathbb{Z}_3 \times \mathbb{Z}_6, \mathbb{Z}_4 \times \mathbb{Z}_4, 
\mathbb{Z}_6 \times \mathbb{Z}_6
\end{array}
$ \\
\hline II  & 1 & 2 & $\kappa_{122}=1, \hat{\kappa}_{111}=-1$ &  $
\begin{array}{c}
\mathbb{Z}_{8-I}, \mathbb{Z}_{12-I}, \mathbb{Z}_2 \times \mathbb{Z}_{6^{\prime}}, \mathbb{Z}_3 \times \mathbb{Z}_3, \\
\mathbb{Z}_3 \times \mathbb{Z}_6, \mathbb{Z}_4 \times \mathbb{Z}_4,
\mathbb{Z}_6 \times \mathbb{Z}_6
\end{array}
$ \\
\hline III  & 1 & 4 & $
\begin{array}{c}
\kappa_{123}=1, \kappa_{144}=-1, \\
\hat{\kappa}_{111}=-1
\end{array}
$  & $\mathbb{Z}_{6-I}$ \\
\hline IV & 3 & 2 & $
\begin{array}{c}
\kappa_{122}=1, \hat{\kappa}_{1 \alpha \alpha}=-1, \\
\alpha \in\{1,2,3\}
\end{array}
 $  & $\mathbb{Z}_{6-I}$ \\
\hline V  & 3 & 6 & $
\begin{array}{c}
\kappa_{123}=1, \kappa_{456}=-2, \\
\kappa_{144}=\kappa_{255}=\kappa_{366}=-2, \\
\hat{\kappa}_{111}=\hat{\kappa}_{222}=\hat{\kappa}_{333}=-1, \\
\hat{\kappa}_{423}=\hat{\kappa}_{513}=\hat{\kappa}_{612}=1,
\end{array}
$  & $\mathbb{Z}_3$ \\
\hline
\end{tabular}
\caption{Selected toroidal orbifold models with $h^{2,1}=0$. The columns indicate their respective Hodge numbers, triple intersection numbers and the various quotients from which they can arise. A hat is used to denote mixed even/odd intersection numbers, which do not enter the F-term scalar potential. Subscripts and primes are used to denote different realizations of the same group. Table adapted from \cite{Flauger:2008ad}.}
\label{tab:mod}
\end{table}

Toroidal orbifolds obtained by quotienting with respect to an abelian group and obeying the Calabi-Yau condition have been classified in \cite{Erler:1992ki,Reffert:2006du,Reffert:2007im}. A partial classification of their orientifold counterparts, which break susy to $\mathcal{N}=1$, was discussed in \cite{Flauger:2008ad}. In Table \ref{tab:mod}, we list all of the examples from \cite{Flauger:2008ad} with $h^{2,1}=0$, grouped into five classes. Although some of them admit multiple realisations, models in each class are characterised by the same topological and geometric data (cohomologies, triple intersection numbers) and give rise to identical low-energy EFTs in the untwisted sector. In particular, class I contains the original, $\mathbb{T}^6/ \mathbb{Z}_3 \times \mathbb{Z}_3$ realisation of DGKT, which was already shown to obey the holographic constraint. Moreover, models II and IV have the same triple intersection numbers for 2-cycles odd under the orientifold involution, and give rise to the same kinetic terms and potential for the scalar moduli. Therefore, it will be sufficient to check the constraint in one of the two cases only. In the remainder of the paper, we will explicitly check the validity of the holographic constraints for models II,III and V of the above table. This will establish the compatibility of models I-V with the constraint \eqref{eq:ctr}. As we will see, the difficulty in the computations increases significantly in scenarios with more moduli and non-canonical K\"ahler metric, such as models III and V.

\section{Explicit models}\label{sc:mods}
In this section we consider explicit realisations of the DGKT vacua described in section \ref{ssc:dgkt}, corresponding to all of the examples presented in Table \ref{tab:mod}. Since $h^{2,1}=0$ for all the examples, the chiral superfields appearing in the untwisted sector are given by the K\"ahler moduli and the axio-dilaton,
\begin{equation}
    t_a=b_a+i v_a \qquad \qquad N=\xi+i z, \qquad \qquad a=1,...,h^{1,1}_-.
\end{equation}
Let the vectors
$\delta\boldsymbol\phi_a$ and $\delta\boldsymbol\phi_s$ denote the axion and saxion fluctuations around the vacuum in these coordinates.
At a supersymmetric vacuum, the kinetic terms \eqref{eq:kt} are written with the same real symmetric matrix $G_*$ in the two sectors
\begin{equation}
{\cal L}_{\rm kin}^{\rm Lor}
=-\frac12(\partial\delta\boldsymbol\phi_a)^TG_*(\partial\delta\boldsymbol\phi_a)
 -\frac12(\partial\delta\boldsymbol\phi_s)^TG_*(\partial\delta\boldsymbol\phi_s),
\end{equation}
where the star symbol denotes quantities evaluated at the vacuum.
The kinetic terms in the vacuum can be abstractly diagonalised with the matrix $U_K$ as
\begin{equation}
U_KG_*U_K^T=K_D,
\end{equation}
where $K_D$ is diagonal and positive definite. Therefore, its square root $\sqrt{K}_D$ is a well defined matrix. Let us now go on to describe the structure of the scalar potential. Again, the Hessian matrix for axions and saxions factorises into two blocks, which we denote as $H_a$ and $H_s$ respectively. The corresponding canonically normalized mass matrices are given by
\begin{equation}
M'_a= \Ra^2 K_D^{-1/2}U_KH_aU_K^TK_D^{-1/2},
\qquad
M'_s= \Ra^2 K_D^{-1/2}U_KH_sU_K^TK_D^{-1/2}.
\end{equation}
Because of supersymmetry, they can be simultaneously diagonalised by a single matrix $U_M$, as 
\begin{equation}
U_MM'_aU_M^T=M_{D,a},\qquad
U_MM'_sU_M^T=M_{D,s}.
\end{equation}
Therefore, the complete map from mass eigenstates to the original
fluctuations is given by
\begin{equation}\label{eq:cob}
\delta\boldsymbol\phi_a=T_{\rm field}\boldsymbol a,\qquad
\delta\boldsymbol\phi_s=T_{\rm field}\boldsymbol \varphi,\qquad
T_{\rm field}=U_K^TK_D^{-1/2}U_M^T,
\end{equation} 
where
\begin{equation}
    \boldsymbol a = (a_i,a_0) \qquad  \boldsymbol \varphi = (\varphi_i,\varphi_0) \qquad i=1,...,h^{1,1}_-.
\end{equation}
The ``singlets'' $a_0,\varphi_0$ denote the scalars of dimension $\Delta=11,10$, while the ``vectors'' $a_i,\varphi_i$ denote those of dimension $\Delta=5,6$.

Although the procedure outlined above is conceptually straightforward, carrying it out explicitly becomes increasingly challenging as the number of moduli grows. In particular, the computation of cubic couplings requires the explicit mass eigenvectors, rather than just the eigenvalues, since the field transformation $T_{\rm field}$ enters three times in Eqs.~\eqref{eq:cdef}-\eqref{eq:ddef}. Moreover, the kinetic and mass matrices depend non-trivially on the fluxes and intersection numbers, making a direct symbolic diagonalisation impractical in more complicated examples. To overcome these difficulties, we exploit the particular structure of the K\"ahler potential to construct suitable intermediate bases. For instance, in Model V we reorganise the six K\"ahler moduli into a symmetric matrix and use its Cholesky decomposition to obtain an explicit diagonalisation of the kinetic matrix. The details of these transformations are reported in Appendix~\ref{app:mod}.

Since the spectrum is always given by Eq.~\eqref{eq:deltadgkt}, there are only two possible extremal arrangements (with multiplicity), corresponding to $(10,5,5)$ and $(11,6,5)$. As explained in Section~\ref{ssc:cctr}, supersymmetry relates the corresponding three-point functions through Eq.~\eqref{eq:csusy}. It is therefore sufficient to verify the holographic constraint for the $(10,5,5)$ arrangement.

In the former case, the extremal cubic couplings can be written as
\begin{equation}\label{eq:cdIII}
    c'_{\varphi_0 a_i a_j} = c_{\varphi_0 a_i a_j} + \frac{70 - 10 - 10}{2} d_{\varphi_0 a_i a_j}.
\end{equation}
To calculate the $d_{\varphi_0 a_i a_j}$'s and $c_{\varphi_0 a_i a_j}$'s, one should first compute them in the original coordinates by taking derivatives of the K\"ahler metric and the scalar potential (evaluated at the vacuum), and later express them in terms of diagonalised fields through the change of basis given in \eqref{eq:cob}. In formulas, this can be written as
\begin{equation}\label{eq:cdef}
d_{\varphi_0 a_i a_j}
=(T_{\rm field})^A{}_0
(T_{\rm field})^B{}_i(T_{\rm field})^C{}_j
\left.\partial_AG_{BC}\right|_*,
\end{equation}
and
\begin{equation}\label{eq:ddef}
c_{\varphi_0 a_i a_j}
= \Ra^2 (T_{\rm field})^A{}_0
(T_{\rm field})^B{}_i(T_{\rm field})^C{}_j
\left.\partial_A\partial_B\partial_C V \right|_*.
\end{equation}
Therefore, once we are able to perform the diagonalisation \eqref{eq:cob}, the cubic couplings can be straightforwardly obtained from Eqs.~\eqref{eq:cdef}-\eqref{eq:ddef}. In the rest of this section we will do this explicitly for all the models in Table \ref{tab:mod}, although some of the details will be relegated to Appendix~\ref{app:mod}.

\subsection{Models II,IV}

Models II and IV have the same untwisted sector with $h^{1,1}_-=2$, and can therefore be analysed together. Moreover, they can be seen as the original $\mathbb{T}^6/\mathbb{Z}_3 \times \mathbb{Z}_3 $ orbifold (model I) with two K\"ahler moduli identified, $t_2=t_3$. Although one might argue that its consistency with \eqref{eq:ctr} already follows from that of the original DGKT orbifold, it is instructive to verify it explicitly.

The volume determines the K\"ahler potential as
\begin{equation}
{\cal V}=\frac12v_1v_2^2,\qquad \qquad
K=-\log(8{\cal V})-4\log(2z),
\end{equation}
while the superpotential is given by
\begin{equation}
W=e_0+e_1t_1+e_2t_2-\frac12m_0t_1t_2^2-2pN.
\end{equation}
As shown in Appendix \ref{ssc:mII}, the F-term equations are solved by
\begin{equation}
\begin{gathered}
b_{1*}=b_{2*}=0,\qquad
\xi_*=\frac{e_0}{2p},\qquad
q_1\equiv v_{1*}
=-\frac{5e_2}{3m_0q_2}, \qquad
q_2\equiv v_{2*}
=\sqrt{-\frac{10e_1}{3m_0}}, \\
{\cal V}_*=\frac12q_1q_2^2,\qquad\qquad 
z_*=-\frac{4m_0{\cal V}_*}{5p},
\end{gathered}
\end{equation}
and require ${\rm{sgn}} (m_0 e_i) <0,\, m_0 p <0$. At the minimum,
\begin{equation}
W_*=\frac{4im_0{\cal V}_*}{5},\qquad
V_{*}=-\frac{75p^4}{1024m_0^2{\cal V}_*^3},\qquad
R_{\rm{AdS}}^2=\frac{2048m_0^2{\cal V}_*^3}{25p^4}.
\end{equation}
The kinetic matrix is already diagonal, and can be written as
\begin{equation}
G(v,z)=\operatorname{diag}\left(\frac{1}{v_1^2},\frac{2}{v_2^2},\frac{4}{z^2}\right),
\qquad
G_*=\operatorname{diag}\left(\frac{1}{q_1^2},\frac{2}{q_2^2},\frac{4}{z_*^2}\right).
\end{equation}
Therefore, it is only necessary to rescale the fields in order to ensure canonical normalisation, and then to diagonalise the mass matrix. All the steps are reported in Appendix \ref{ssc:mII}, and result in the known spectrum
\begin{equation}\label{eq:modIIs}
    m^2_a R^2_{\rm{AdS}}=(10,10,88) \qquad \quad \qquad  m^2_s R^2_{\rm{AdS}}=(18,18,70),
\end{equation}
corresponding to the conformal dimensions
\begin{equation}
    \Delta_a=(5,5,11) \qquad \quad \qquad \Delta_s=(6,6,10).
\end{equation}

For the $(10,5,5)$ cubic arrangement, the couplings in a suitable mass eigenstate basis can be obtained as
\begin{equation}
d_{\varphi_0 a_i a_j}=
\begin{pmatrix}
\frac{4}{\sqrt{13}}&0\\
0&-\frac{8}{13\sqrt{13}}
\end{pmatrix},
\qquad
c_{\varphi_0 a_i a_j}=
\begin{pmatrix}
-\frac{100}{\sqrt{13}}&0\\
0& \frac{200}{13\sqrt{13}}
\end{pmatrix},
\end{equation}
and
\begin{equation}
c_{\varphi_0 a_i a_j}+25d_{\varphi_0 a_i a_j}=0
\end{equation}
as expected.

\subsection{Model III}
Model III provides the first genuinely non-trivial example of the diagonalisation procedure described above. Unlike models II and IV, the kinetic matrix is non-diagonal already at the vacuum and depends non-trivially on the flux parameters. Extracting explicit canonically normalised mass eigenstates, rather than merely the eigenvalues, requires a suitable choice of intermediate basis. For this reason model III provides an important test of whether the cancellation observed in the simpler examples survives once the individual cubic couplings depend in a more complicated way on the compactification parameters.

The volume and K\"ahler potential are given by
\begin{equation}
{\cal V}=v_1(v_2v_3-\frac{1}{2}v_4^2) \equiv v_1D(v),\qquad \qquad
K=-\log(8{\cal V})-4\log(2z),
\end{equation}
\begin{equation}
W=e_0+e_1t_1+e_2t_2+e_3t_3+e_4t_4
-m_0t_1\left(t_2t_3-\frac12t_4^2\right)-2pN.
\end{equation}
As usual, the F-term equations are given in Appendix \ref{ssc:mIII}, and result in the supersymmetric minimum
\begin{equation}\label{eq:modIIIvev}
\begin{gathered}
q_1\equiv v_{1*}
=\sqrt{-\frac{5\Delta_e}{3m_0e_1}},\qquad
q_2\equiv v_{2*}=-\frac{5e_3}{3m_0q_1}, \qquad
q_3\equiv v_{3*}=-\frac{5e_2}{3m_0q_1},\\
q_4\equiv v_{4*}=\frac{5e_4}{3m_0q_1},\qquad
z_*=-\frac{4m_0{\cal V}_*}{5p},\qquad
b_{a*}=0,\qquad \xi_*=\frac{e_0}{2p},\\ \qquad
D_*=q_2q_3-\frac12q_4^2=-\frac{5e_1}{3m_0},\qquad \qquad
{\cal V}_*=q_1D_*.
\end{gathered}
\end{equation}
where we have defined
\begin{equation}
\Delta_e=e_2e_3-\frac12e_4^2.
\end{equation}
Similarly to the previous case, the fluxes need to satisfy various inequalities in order to give positive volumes, satisfy the tadpole constraint and for the square root factors to be real. From now on, we assume this is the case. At the minimum, 
\begin{equation}
W_*=\frac{4im_0{\cal V}_*}{5},\qquad
V_{*}=-\frac{75p^4}{1024m_0^2{\cal V}_*^3},\qquad
\Ra^2=\frac{2048m_0^2{\cal V}_*^3}{25p^4}.
\end{equation}
As usual, quantities evaluated at the minimum are denoted by a star, and $q_a\equiv v_{a*}$. In this case, the kinetic term matrix
\begin{equation}
G(v,z)=
\begin{pmatrix}\label{eq:modIIIkin}
\dfrac1{v_1^2}&0&0&0&0\\
0&\dfrac{v_3^2}{D^2}&\dfrac{v_4^2}{2D^2}
&-\dfrac{v_3v_4}{D^2}&0\\
0&\dfrac{v_4^2}{2D^2}&\dfrac{v_2^2}{D^2}
&-\dfrac{v_2v_4}{D^2}&0\\
0&-\dfrac{v_3v_4}{D^2}&-\dfrac{v_2v_4}{D^2}
&\dfrac{v_2v_3+v_4^2/2}{D^2}&0\\
0&0&0&0&\dfrac4{z^2}
\end{pmatrix},
\qquad
G_*=G(q,z_*).
\end{equation}
is non-diagonal, even when evaluated at the minimum. It can be diagonalised simultaneously with the mass matrix as described at the beginning of this section, with the details provided in Appendix \ref{ssc:mIII}.

The $(10,5,5)$ cubic couplings in a convenient mass eigenstate basis then become
\begin{equation}
 d_{\varphi_0 a_i a_j}
 =
 \operatorname{diag}\left(
 \frac{4}{\sqrt{13}},
 \frac{4}{\sqrt{13}},
 \frac{4}{\sqrt{13}},
 -\frac{8}{13\sqrt{13}}
 \right),
 \end{equation}
 and
 \begin{equation}
 c_{\varphi_0 a_i a_j}
 =
 \operatorname{diag}\left(
 -\frac{100}{\sqrt{13}},
 -\frac{100}{\sqrt{13}},
 -\frac{100}{\sqrt{13}},
 \frac{200}{13\sqrt{13}}
 \right).
\label{eq:modelIII-couplings}
\end{equation}
Again,
\begin{equation}
c_{\varphi_0 a_i a_j}+25d_{\varphi_0 a_i a_j}=0,
\end{equation}
as expected.

\subsection{Model V}

Finally, we turn to Model V, which is the most technically involved example in our analysis. With six K\"ahler moduli, the kinetic and mass matrices become considerably more complicated, making a direct symbolic diagonalisation impractical. \footnote{In practice, a brute-force symbolic diagonalisation of the flux-dependent kinetic and mass matrices using standard Mathematica routines
does not yield tractable expressions for the eigenvectors. The analytic construction described below circumvents this difficulty.} As we will see, the determinant structure of
the volume provides a useful simplification.

The kinetic terms are determined by the usual K\"ahler potential
\begin{equation}
K=-\log(8\mathcal{V})-4\log(2z),
\end{equation}
where now the volume takes the more complicated form
\begin{equation}
{\cal V}=
v_1v_2v_3-v_1v_4^2-v_2v_5^2-v_3v_6^2-2v_4v_5v_6.
\end{equation}
In this case, it is convenient to repackage the fields and fluxes into the matrices
\begin{equation}
 {\bf{t}}(t)=
\begin{pmatrix}
t_1&t_6&t_5\\
t_6&t_2&-t_4\\
t_5&-t_4&t_3
\end{pmatrix},
\qquad
M(v)=\operatorname{Im}{ {\bf{t}}}(t),
\qquad
{\bf{e}}=
\begin{pmatrix}
e_1&e_6/2&e_5/2\\
e_6/2&e_2&-e_4/2\\
e_5/2&-e_4/2&e_3
\end{pmatrix}.
\end{equation}
Then, one can rewrite the K\"ahler potential more conveniently as a function of ${\cal V}=\det M$. The superpotential also admits the clean expression
\begin{equation}
W=e_0+{\rm{Tr}}( {\bf{e}}{\bf{t}})-m_0\det{{\bf{t}}}-2pN.
\end{equation}

The determinant structure of the volume provides a useful simplification in the diagonalisation procedure. Indeed, introducing the symmetric generators $E_a$ defined in Appendix~\ref{ssc:modV}, the K\"ahler metric can be written as
\begin{equation}
(G_K)_{ab}
=
\rm{Tr}\left(M^{-1}E_a M^{-1}E_b\right).
\end{equation}
At the vacuum, the positive-definite matrix $q=M|_*$ admits a Cholesky decomposition $q=LL^T$. Denoting fluctuations around the vacuum by $\delta M\equiv M-q$, and defining $\delta\widehat M=L^{-1}\delta M L^{-T}$, the kinetic quadratic form becomes
\begin{equation}
    \mathrm{Tr}\left(q^{-1}\delta M q^{-1}\delta M\right)
    =
    \mathrm{Tr}\left(\delta\widehat M^2\right).
\end{equation}
This allows us to construct a basis in which the flux-dependent kinetic matrix is diagonal algebraically, without directly diagonalising it. In particular, the latter reduces to a diagonal matrix with constant entries, $K_D=\operatorname{diag}(1,1,1,2,2,2,4)$, where the last entry corresponds to the dilaton. Canonical normalisation is then achieved by the simple rescaling $K_D^{-1/2}$, after which the remaining mass matrix can be diagonalised by a flux-independent orthogonal transformation, as detailed in Appendix~\ref{ssc:modV}.

By inspecting the F-terms, the supersymmetric vacuum is located at
\begin{equation}\label{eq:modVvev}
\begin{gathered}
v_{1*}\equiv q_1=\bar v\left(e_2e_3-\frac{e_4^2}{4}\right),\qquad
v_{2*}\equiv q_2=\bar v\left(e_1e_3-\frac{e_5^2}{4}\right),\\
v_{3*}\equiv q_3=\bar v\left(e_1e_2-\frac{e_6^2}{4}\right), \qquad
v_{4*}\equiv q_4=-\bar v\left(\frac{e_1e_4}{2}+\frac{e_5e_6}{4}\right),\\ 
v_{5*}\equiv q_5 =-\bar v\left(\frac{e_2e_5}{2}+\frac{e_4e_6}{4}\right),\qquad
v_{6*}\equiv q_6 =-\bar v\left(\frac{e_3e_6}{2}+\frac{e_4e_5}{4}\right),\\
z_*=-\frac{4m_0D}{5p},\qquad b_{a*}=0,\qquad \xi_*=\frac{e_0}{2p},
\end{gathered}
\end{equation}
where we have also defined
\begin{equation}
\Delta_E=\det {\bf{e}}
=e_1e_2e_3-\frac{e_1e_4^2+e_2e_5^2+e_3e_6^2+e_4e_5e_6}{4},
\qquad
\bar v=\frac{1}{\sqrt{-\frac{3m_0}{5}\Delta_E}},
\end{equation}
and again the fluxes have to satisfy various inequalities that we assume to be true. At the minimum, the potential and superpotential take the values
\begin{equation}
W_*=\frac{4im_0D}{5},\qquad
V_{*}=-\frac{75p^4}{1024m_0^2D^3},\qquad
\Ra^2=\frac{2048m_0^2D^3}{25p^4},
\end{equation}
where $D=\det M(v_*)$. In this case, the expression for the kinetic matrices $G(v,z)$ and $G_*(v,z)$ are significantly more complicated, and reported only in the Appendix \ref{ssc:modV}.

Upon performing the same diagonalization procedure as in the other cases, we can write down the results for the extremal $(10,5,5)$ cubic couplings. In an appropriate basis of mass eigenstates,
\begin{equation}
 d_{\varphi_0 a_i a_j}
 =
 \operatorname{diag}\left(
 \frac{4}{\sqrt{13}},
 \frac{4}{\sqrt{13}},
 \frac{4}{\sqrt{13}},
 \frac{4}{\sqrt{13}},
 \frac{4}{\sqrt{13}},
 -\frac{8}{13\sqrt{13}}
 \right),
 \end{equation}
 and
 \begin{equation}
 c_{\varphi_0 a_i a_j}
 =
 \operatorname{diag}\left(
 -\frac{100}{\sqrt{13}},
 -\frac{100}{\sqrt{13}},
 -\frac{100}{\sqrt{13}},
 -\frac{100}{\sqrt{13}},
 -\frac{100}{\sqrt{13}},
 \frac{200}{13\sqrt{13}}
 \right),
\label{eq:modelV-couplings}
\end{equation}
which take the same form as the previous expressions and yet again verify
\begin{equation}
c_{\varphi_0 a_i a_j}+25d_{\varphi_0 a_i a_j}=0.
\end{equation}

\section{Discussion}\label{sc:conc}

The main result of this note is to explicitly verify the validity of the holographic constraint of \cite{Bobev:2025yxp} for DGKT-type vacua in massive type IIA String Theory. In particular, we focused on the classification of toroidal abelian orientifolds presented in \cite{Flauger:2008ad} (see Table \ref{tab:mod}), and showed that extremal 3-point couplings always vanish in all the examples with $h^{2,1}=0$. In all of the models we considered, there are two possible choices of extremal arrangements which must satisfy the constraint, whose conformal dimensions are given by the triples $(10,5,5)$ and $(11,6,5)$. An additional ingredient in our analysis is the use of $3d$ $\mathcal N=1$ superconformal symmetry. The superspace relations derived in Appendix~\ref{app:3d1} explain the ratios between non-extremal three-point coefficients previously obtained from explicit holographic
calculations in~\cite{Bobev:2025yxp}. Moreover, they relate the two extremal arrangements $(10,5,5)$ and $(11,6,5)$,
reducing the holographic consistency check to a single
independent class of correlators. This simplification
was subsequently exploited in the general analysis
of~\cite{Revello:2026yla}. By contrast, when $h^{2,1}\neq0$ and complex-structure moduli are present, violations of the cubic coupling constraint can occur \cite{Revello:2026eqp}, with the general structure analysed subsequently in \cite{Revello:2026yla}.

The explicit verification of the constraint is considerably more involved than the simplicity of the final result might suggest. Direct computations can become more challenging in models with more moduli than the original $\mathbb{T}^6/\mathbb{Z}_3 \times \mathbb{Z}_3$ example as one needs not only the eigenvalues but also canonically normalised eigenvectors. For instance, the kinetic term matrix for the moduli, obtained by differentiating the K\"ahler potential, is in general non-diagonal, and must be explicitly diagonalised with a matrix $U_K$, which can depend on fluxes and triple intersection numbers. It is therefore even more surprising that the cubic scalar and derivative couplings - denoted as $c_{ijk}$'s and $d_{ijk}$'s in the text - always combine in such a way as to ensure an exact cancellation in these complicated cases. In practice, we only had to verify this explicitly for the $(10,5,5)$ extremal arrangement, and found that the values for the individual contributions to the cubic couplings also seem to exhibit a very clear pattern. The $c_{ijk}$'s and $d_{ijk}$'s for the $(10,5,5)$ extremal arrangement are given by
\begin{equation}\label{eq:dfin}
d_{\varphi_0 a_i a_j}
=
\operatorname{diag}\Big(
\underbrace{\frac{4}{\sqrt{13}},\ldots,\frac{4}{\sqrt{13}}}_{h^{1,1}_- -1\ \text{entries}},
-\frac{8}{13\sqrt{13}}
\Big),
\end{equation}
and
\begin{equation}\label{eq:cfin}
c_{\varphi_0 a_i a_j}
=
\operatorname{diag}\Big(
\underbrace{-\frac{100}{\sqrt{13}},\ldots,-\frac{100}{\sqrt{13}}}_{h^{1,1}_- -1\ \text{entries}},
\frac{200}{13\sqrt{13}}
\Big).
\end{equation}
Notice how, modulo a change of basis, the above equations are also compatible with the expressions found for the $\mathbb{T}^6/\mathbb{Z}_3 \times \mathbb{Z}_3$ example in \cite{Bobev:2025yxp}, which corresponds to model I in the classification of \cite{Flauger:2008ad}.\footnote{The matrices in Eqs. (23)-(25) of \cite{Bobev:2025yxp} have eigenvalues equal to those of our \eqref{eq:dfin}-\eqref{eq:cfin} for $h^{1,1}_-=3$. Therefore, they are equivalent up to an $O(3)$ rotation.} The structure of these equations already seems to point towards a geometric explanation of the universality. The $h^{1,1}_- -1$ identical entries correspond to K\"ahler deformations which leave the volume invariant, and the remaining one arises from the overall volume which mixes with the dilaton. This pattern emerged from the explicit diagonalisation of the examples studied here, before a general construction of the corresponding eigenvectors was available. It both motivated the search for a model-independent proof and suggested the explicit decomposition of the K\"ahler moduli into volume-preserving and volume-changing directions that is used in the subsequent general analysis of \cite{Revello:2026yla}. In this sense, the explicit examples did not merely provide checks of the general result, but supplied an essential clue for its derivation.

The situation is highly reminiscent of the one encountered in \cite{Apers:2022tfm}, where it was shown that the low-lying spectrum of DGKT-type vacua on a general Calabi-Yau orientifold is in fact universal, unlike other intermediate quantities (such as the scalar potential) which depend explicitly on fluxes and the internal geometry. The universal structure uncovered in these examples suggests that the cancellations do not depend on the details of the
compactification geometry. This expectation was subsequently
confirmed in~\cite{Revello:2026yla}, where the cancellation was proven for arbitrary Calabi--Yau orientifolds with $h^{2,1}=0$, building on the structure of the mass eigenvectors identified in the present work.

\section*{Acknowledgements}
We would like to thank Nikolay Bobev, Joe Conlon, Hynek Paul and Thomas Van Riet for insightful discussions, and in particular Vincent Van Hemelryck for collaboration on the related article \cite{Revello:2026yla}. FR is also grateful to the organizers and participants of \emph{Lotus \& Swamplandia 2026} and of the workshop \emph{Expanding Thoughts on de Sitter}, where part of this work was carried out, and to the ``Espacio La Ricotta'' for hospitality. Moreover, FR would like to especially thank Muthusamy Rajaguru for conversations regarding the use of agentic AI. FR acknowledges support from a junior postdoctoral fellowship of the Fonds Wetenschappelijk Onderzoek (FWO), project number 12A1Q25N.

\paragraph*{AI assistance}
The authors used large language models to assist with
generating and verifying Mathematica notebooks, as well
as with language editing and improving the presentation
of the manuscript. All scientific results and computational
outputs were independently verified by the authors, who
retain full responsibility for the content of this work.

\appendix

\section{Details on the explicit models}\label{app:mod}
\subsection{Models II,IV}\label{ssc:mII}

The F-term equations are
\begin{align}
    &D_N W = -2p+\frac{2i(e_0+e_1t_1+e_2t_2-\frac12m_0t_1t_2^2-2pN)}{z} = 0, \\
    &D_{t_1} W = e_1-\frac12m_0t_2^2+\frac{i(e_0+e_1t_1+e_2t_2-\frac12m_0t_1t_2^2-2pN)}{2v_1} = 0, \\
    &D_{t_2} W = e_2-m_0t_1t_2+\frac{i(e_0+e_1t_1+e_2t_2-\frac12m_0t_1t_2^2-2pN)}{v_2} = 0. 
\end{align}
The kinetic matrix is already diagonal, hence
\begin{equation}
U_K=\operatorname{diag}(q_1,q_2,z_*),\qquad
K_D=\operatorname{diag}\left(1,2,4\right),\qquad
K_D^{-1/2}=\operatorname{diag}\left(1,\frac{1}{\sqrt{2}},\frac{1}{2}\right).
\end{equation}
In the notation of the main text, the mass matrix is diagonalised by
\begin{equation}
U_M=
\begin{pmatrix}
-\sqrt{2/3}&1/\sqrt3&0\\
1/\sqrt{39}&\sqrt{2/39}&2\sqrt{3/13}\\
-2/\sqrt{13}&-2\sqrt{2/13}&1/\sqrt{13}
\end{pmatrix},
\end{equation}
and hence the complete transformation is given by
\begin{equation}
T_{\rm field}=U_K^TK_D^{-1/2}U_M^T
=
\begin{pmatrix}
-q_1\sqrt{2/3}&q_1/\sqrt{39}&-2q_1/\sqrt{13}\\
q_2/\sqrt6&q_2/\sqrt{39}&-2q_2/\sqrt{13}\\
0&z_*\sqrt{3/13}&z_*/(2\sqrt{13})
\end{pmatrix}.
\end{equation}
Thus
\begin{equation}
\begin{pmatrix}b_1\\ b_2\\ \xi-\xi_*\end{pmatrix}
=T_{\rm field}\begin{pmatrix}a_1\\ a_2\\ a_0\end{pmatrix},
\qquad
\begin{pmatrix}v_1-q_1\\ v_2-q_2\\ z-z_*\end{pmatrix}
=T_{\rm field}\begin{pmatrix}\varphi_1\\ \varphi_2\\ \varphi_0\end{pmatrix}.
\end{equation}
The above matrices satisfy the identities
\begin{equation}
U_KG_*U_K^T=K_D,\qquad \qquad
T_{\rm field}^TG_*T_{\rm field}=I_3,
\end{equation}
and
\begin{equation}
\Ra^2 T_{\rm field}^TH_aT_{\rm field}
=\operatorname{diag}(10,10,88),\qquad
\Ra^2 T_{\rm field}^TH_sT_{\rm field}
=\operatorname{diag}(18,18,70).
\end{equation}
resulting in the spectrum \eqref{eq:modIIs}.

\subsection{Model III}\label{ssc:mIII}

The F-term equations are given by
\begin{equation}
\begin{gathered}
D_N W=-2p+2i\frac{e_0+e_1t_1+e_2t_2+e_3t_3+e_4t_4
-m_0t_1\left(t_2t_3-\frac12t_4^2\right)-2pN}{z}=0,\\
D_{t_1}W=e_1-m_0\left(t_2t_3-\frac12t_4^2\right)+i\frac{e_0+e_1t_1+e_2t_2+e_3t_3+e_4t_4
-m_0t_1\left(t_2t_3-\frac12t_4^2\right)-2pN}{2v_1}=0,\\
D_{t_2} W=e_2-m_0t_1t_3+i v_3\frac{e_0+e_1t_1+e_2t_2+e_3t_3+e_4t_4
-m_0t_1\left(t_2t_3-\frac12t_4^2\right)-2pN}{2D(v)}=0,\\
D_{t_3} W=e_3-m_0t_1t_2+iv_2\frac{e_0+e_1t_1+e_2t_2+e_3t_3+e_4t_4
-m_0t_1\left(t_2t_3-\frac12t_4^2\right)-2pN}{2D(v)}=0,\\
D_{t_4} W=e_4+m_0t_1t_4-iv_4\frac{e_0+e_1t_1+e_2t_2+e_3t_3+e_4t_4
-m_0t_1\left(t_2t_3-\frac12t_4^2\right)-2pN}{2D(v)}=0,\\
\end{gathered}
\end{equation}
and they are solved by \eqref{eq:modIIIvev}. The kinetic terms in the vacuum are given by the matrix \eqref{eq:modIIIkin}. Defining
\begin{equation}
F=
\begin{pmatrix}
q_1&0&0&0&0\\
0&q_2&0&0&0\\
0&\dfrac{q_4^2}{2q_2}&\dfrac{D_*}{q_2}
&\dfrac{q_4\sqrt{D_*}}{q_2}&0\\
0&q_4&0&\sqrt{D_*}&0\\
0&0&0&0&z_*
\end{pmatrix},
\end{equation}
they can be diagonalised as
\begin{equation}
U_K=F^T,\qquad
K_D=\operatorname{diag}\left(1,1,1,1,4\right),\qquad
K_D^{-1/2}=\operatorname{diag}\left(1,1,1,1,\frac12\right).
\end{equation}
In this basis, the mass matrix is diagonalised by
\begin{equation}
U_M=
\begin{pmatrix}
-1/\sqrt2&0&1/\sqrt2&0&0\\
-1/\sqrt6&\sqrt{2/3}&-1/\sqrt6&0&0\\
0&0&0&1&0\\
1/\sqrt{39}&1/\sqrt{39}&1/\sqrt{39}&0&2\sqrt{3/13}\\
-2/\sqrt{13}&-2/\sqrt{13}&-2/\sqrt{13}&0&1/\sqrt{13}
\end{pmatrix}.
\end{equation}
Define
\begin{equation}
A=\frac{q_4^2}{2q_2},\qquad
B=\frac{D_*}{q_2},\qquad
C=\frac{q_4\sqrt{D_*}}{q_2}.
\end{equation}
The overall field transformation is
\begin{equation}
T_{\rm field}=U_K^TK_D^{-1/2}U_M^T
=
\begin{pmatrix}
-q_1/\sqrt2&-q_1/\sqrt6&0&q_1/\sqrt{39}&-2q_1/\sqrt{13}\\
0&q_2\sqrt{2/3}&0&q_2/\sqrt{39}&-2q_2/\sqrt{13}\\
B/\sqrt2&A\sqrt{2/3}-B/\sqrt6&C&(A+B)/\sqrt{39}
&-2(A+B)/\sqrt{13}\\
0&q_4\sqrt{2/3}&\sqrt{D_*}&q_4/\sqrt{39}
&-2q_4/\sqrt{13}\\
0&0&0&z_*\sqrt{3/13}&z_*/(2\sqrt{13})
\end{pmatrix},
\end{equation}
acting as
\begin{equation}
\delta\boldsymbol\phi_a= T_{\rm field}
\begin{pmatrix}a_1\\ a_2\\ a_3\\ a_4\\ a_0\end{pmatrix},
\qquad
\delta\boldsymbol\phi_s= T_{\rm field}
\begin{pmatrix} \varphi_1\\ \varphi_2\\ \varphi_3\\ \varphi_4\\ \varphi_0\end{pmatrix}.
\end{equation}
The above matrices obey the identities
\begin{equation}
U_KG_*U_K^T=K_D,\qquad
T_{\rm field}^TG_*T_{\rm field}=I_5,
\end{equation}
resulting in the expected spectrum
\begin{equation}
\Ra^2T_{\rm field}^T H_a T_{\rm field}
=\operatorname{diag}(10,10,10,10,88),\qquad
\Ra^2T_{\rm field}^T H_s T_{\rm field}
=\operatorname{diag}(18,18,18,18,70).
\end{equation}

\subsection{Model V}\label{ssc:modV}
In analogy with the expression in the main text, let us define
\begin{equation}
{\bf{q}}=M(q)=
\begin{pmatrix}
q_1&q_6&q_5\\
q_6&q_2&-q_4\\
q_5&-q_4&q_3
\end{pmatrix},
\qquad
D=\det {\bf{q}}.
\end{equation}
The full F-term equations $D_{t_a}W=0, D_N W=0$ can then be written in matrix form as
\begin{equation}
0=-2p+\frac{2iW}{z},\qquad
0={\bf{e}}-m_0 {\rm adj}({\bf{t}})+\frac{iW}{2}M^{-1}.
\end{equation}
The real component immediately imposes $b_{a*}=0$ and $\xi_*=e_0/(2p)$. Then,
the remaining equations reduce to
\begin{equation}
E=-\frac{3m_0}{5} {\rm adj} {\bf{q}},\qquad
pz_*=-\frac45m_0D,\qquad
W_*=\frac{4im_0D}{5}.
\end{equation}
The complete solution is given in the main text, in \eqref{eq:modVvev}. To diagonalise the kinetic matrix, it is useful to define  the six symmetric generators
\begin{equation}
\begin{gathered}
E_1=\begin{pmatrix}1&0&0\\0&0&0\\0&0&0\end{pmatrix},\qquad
E_2=\begin{pmatrix}0&0&0\\0&1&0\\0&0&0\end{pmatrix},\qquad
E_3=\begin{pmatrix}0&0&0\\0&0&0\\0&0&1\end{pmatrix},\\
E_4=\begin{pmatrix}0&0&0\\0&0&-1\\0&-1&0\end{pmatrix},\qquad
E_5=\begin{pmatrix}0&0&1\\0&0&0\\1&0&0\end{pmatrix},\qquad
E_6=\begin{pmatrix}0&1&0\\1&0&0\\0&0&0\end{pmatrix}.
\end{gathered}
\end{equation}
Then, the kinetic matrix is expressed as
\begin{equation}
G(v,z)=\operatorname{diag}\left(G_K(M),\frac4{z^2}\right),
\qquad
G_*=\operatorname{diag}\left(G_K({\bf{q}}),\frac4{z_*^2}\right), 
\end{equation}
with
\begin{equation}
(G_K)_{ab}(M)={\rm Tr}(M^{-1}E_aM^{-1}E_b).
\end{equation}
One can perform the Cholesky decomposition ${\bf{q}}=L L^T$, with $L$ of the schematic form
\begin{equation}
L=
\begin{pmatrix}
a&0&0\\
b&c&0\\
d&r&f
\end{pmatrix},
\end{equation}
and with matrix elements given by
\begin{equation}
a=\sqrt{q_1},\qquad
b=\frac{q_6}{\sqrt{q_1}},\qquad
c=\sqrt{\frac{q_1q_2-q_6^2}{q_1}},
\end{equation}
\begin{equation}
d=\frac{q_5}{\sqrt{q_1}},\qquad
r=-\frac{q_1q_4+q_5q_6}{\sqrt{q_1(q_1q_2-q_6^2)}},\qquad
f=\sqrt{\frac{D}{q_1q_2-q_6^2}}.
\end{equation}
This decomposition provides a crucial simplification, rather than directly diagonalising the full flux-dependent $7\times7$ kinetic matrix, it allows us to construct a canonical basis algebraically.

The kinetic terms can be diagonalised by
\begin{equation}
F=
\begin{pmatrix}
a^2&0&0&0&0&0&0\\
b^2&c^2&0&0&0&2bc&0\\
d^2&r^2&f^2&-2rf&2df&2dr&0\\
-bd&-cr&0&cf&-bf&-(br+cd)&0\\
ad&0&0&0&af&ar&0\\
ab&0&0&0&0&ac&0\\
0&0&0&0&0&0&z_*
\end{pmatrix},
\end{equation}
and our usual coordinate transformations are defined as
\begin{equation}
U_K=F^T,\qquad
K_D=\operatorname{diag}\left(1,1,1,2,2,2,4\right),
\qquad
K_D^{-1/2}=\operatorname{diag}
\left(1,1,1,\frac1{\sqrt2},\frac1{\sqrt2},\frac1{\sqrt2},\frac12\right).
\end{equation}
The mass matrix is instead diagonalised by
\begin{equation}
U_M=
\begin{pmatrix}
-1/\sqrt2&0&1/\sqrt2&0&0&0&0\\
-1/\sqrt6&\sqrt{2/3}&-1/\sqrt6&0&0&0&0\\
0&0&0&0&0&1&0\\
0&0&0&0&1&0&0\\
0&0&0&1&0&0&0\\
1/\sqrt{39}&1/\sqrt{39}&1/\sqrt{39}&0&0&0&2\sqrt{3/13}\\
-2/\sqrt{13}&-2/\sqrt{13}&-2/\sqrt{13}&0&0&0&1/\sqrt{13}
\end{pmatrix},
\end{equation}
so that the overall coordinate change is given by
\begin{equation}
T_{\rm field}=U_K^TK_D^{-1/2}U_M^T=FK_D^{-1/2}U_M^T.
\end{equation}
Thus
\begin{equation}
\delta\boldsymbol\phi_a=T_{\rm field}
\begin{pmatrix}
a_1\\ a_2\\ a_3\\a_4\\ a_5\\ a_6\\ a_0
\end{pmatrix},
\qquad
\delta\boldsymbol\phi_s=T_{\rm field}
\begin{pmatrix}
\varphi_1\\ \varphi_2\\ \varphi_3\\ \varphi_4\\ \varphi_5\\ \varphi_6\\ \varphi_0
\end{pmatrix}.
\end{equation}
The diagonalisation matrices satisfy
\begin{equation}
U_KG_*U_K^T=K_D,\qquad
T_{\rm field}^TG_*T_{\rm field}=I_7,
\end{equation}
and they determine the spectrum as
\begin{equation}
\Ra^2T_{\rm field}^TH_aT_{\rm field}
=\operatorname{diag}(10,10,10,10,10,10,88),
\end{equation}
\begin{equation}
\Ra^2 T_{\rm field}^TH_sT_{\rm field}
=\operatorname{diag}(18,18,18,18,18,18,70).
\end{equation}

\section{$3d$ $\mathcal{N}=1$ superspace}\label{app:3d1}
\subsection{Basic formalism}
In this Appendix, we give a brief outline of the superspace formalism for $3d$ $\mathcal{N}=1$ SCFTs, and its use in the computation of two- and three-point functions \cite{Park:1999cw,Buchbinder:2015qsa} (see also \cite{Osborn:1998qu} for analogous applications in $4d$ $\mathcal{N}=1$). More details can be found in the original references, and we adopt the conventions and notation of \cite{Buchbinder:2015qsa}. In particular, we use it to derive relations between 3-point functions containing conformal primaries within the same supersymmetric multiplet, \emph{i.e.} saxions and axions in our language. Similar relations have also been worked out in the context of the 3d $\mathcal{N}=1$ superconformal bootstrap \cite{Rong:2018okz}.

In 3d, superspace coordinates are parametrised by
\begin{equation}
    z^{\mu}=(x^{\mu},\theta^{\alpha}), \quad \quad \quad \mu=0,1,2 \quad \alpha =1,2,
\end{equation}
where the $x^{\mu}$ are spacetime coordinates and $\theta^{\alpha}$ is a real Majorana spinor. Through the 3d gamma matrices, which obey the usual anti-commutation relation $\left\{ \gamma^{\mu},\gamma^{\nu} \right\}= 2 \eta^{\mu \nu}$, one can repackage the spacetime coordinates (or any other vector/tensor) as the matrix
\begin{equation}
    x_{\alpha \beta} = x^{\mu} \left(\gamma_{\mu} \right)_{\alpha \beta}.
\end{equation}
Furthermore, we also define 3d spatial parity $\mathcal{P}$ acting as the transformation
\begin{equation}
    \mathcal{P}: (x^0,x^1,x^2) \rightarrow (x^0,-x^1,x^2),
\end{equation}
which reverses orientation as $\det (\mathcal{P})=-1$. The operators dual to a saxion and an axion sit within a long superconformal multiplet $\Phi$, whose bottom component is a scalar superconformal primary \cite{Cordova:2016emh}. It can be decomposed as\footnote{Not to be confused with the analogous decomposition for a lagrangian field: the components are abstract operators, and $\mathcal{O}_T$ is \emph{not} an auxiliary field.} 
\begin{equation}\label{eq:Phi}
    \Phi= \mathcal{O}_B+ \theta^{\alpha} \psi_{\alpha}+\frac{\theta^2 \mathcal{O}_T}{\sqrt{2 \Delta_B(2\Delta_B-1)}},
\end{equation}
where the conformal primary operators $\mathcal{O}_B$ and $\mathcal{O}_T$ are the bottom and top component of the multiplet. Their respective conformal dimensions $\Delta_B,\Delta_T$ satisfy the relation $\Delta_T=\Delta_B+1$, because the top component can be obtained from the bottom one by acting twice with the supercharge as $\mathcal{O}_T = Q^2 \mathcal{O}_B$. The constants in \eqref{eq:Phi} have been chosen to have normalised 2-pt functions for both $\mathcal{O}_B$ and $\mathcal{O}_T$. Moreover if $\Phi$ has definite spatial parity $P$, the operators  $\mathcal{O}_B$ and $\mathcal{O}_T$ have parities $P$ and $-P$ respectively, as $\mathcal{P}(\theta^2)=-1$. This is exactly what happens for the long multiplets containing scalars dual to a saxion and an axion. In particular, it allows us to interpret multiplets where the saxion is the bottom component as parity even, and those where the axion is the bottom component as parity odd.

\subsection{Two- and three-point functions}
To make the action of supersymmetry manifest, one can define the supersymmetric interval
\begin{equation}
    y_{12}^{\alpha \beta} = (x_1-x_2)^{\alpha \beta} + 2 i \theta_1^{( \alpha} \theta_2^{\beta)}-i (\theta_1-\theta_2)^{\alpha} (\theta_1-\theta_2)^{\beta},
\end{equation}
which is annihilated by the $Q^{\alpha}$. Such an interval is therefore invariant under supersymmetric transformations, but transforms non-trivially under the conformal group. In particular, its norm $y_{12}^2 \equiv -\frac{1}{2} (y_{12})^{\alpha \beta} (y_{12})_{\alpha \beta} $ transforms as
\begin{equation}
    y_{12}^2 \rightarrow \Omega(x_1) \Omega(x_2) y_{12}^2
\end{equation}
under a conformal transformation $\Omega (x)$. Then, as in ordinary CFTs, the two-point function of a superconformal primary $\Phi$ with dimension $\Delta$ is uniquely determined by superconformal invariance, as
\begin{equation}
    \langle \Phi_i (z_1) \Phi_j (z_2) \rangle = \frac{\delta_{ij}}{(y_{12}^2)^{\Delta}}.
\end{equation}
The supersymmetric interval can be expanded as
\begin{equation}\label{eq:exp}
  \frac{1}{(y_{12}^2)^{\Delta}} \big \lvert_{\theta_1^2,\theta_2^2} = 4 \Delta \left(\Delta-\frac{1}{2}\right) \theta_1^2 \theta_2^2  \frac{1}{(x_{12}^2)^{\Delta+1}},
\end{equation}
so that with the normalisation in \eqref{eq:Phi} the two-point functions of the primaries
\begin{equation}
    \langle \mathcal{O}_{B,i} (x_1) \mathcal{O}_{B,j} (x_2) \rangle = \frac{\delta_{ij}}{(x_{12}^2)^{\Delta}} \qquad \quad  \langle \mathcal{O}_{T,i} (x_1) \mathcal{O}_{T,j} (x_2) \rangle = \frac{\delta_{ij}}{(x_{12}^2)^{\Delta+1}}.
\end{equation}

One can also define trilinear structures in the superspace coordinates which transform nicely under the conformal group, namely
\begin{equation}
(\Theta_3)^\alpha
=
(x_{13}^{-1})^{\alpha}{}_{\beta}\,
\theta_{13}^{\beta}
-
( x_{23}^{-1})^{\alpha}{}_{\beta}\,
\theta_{23}^{\beta}, \quad \quad \quad \Theta_3^2 = -\frac{1}{2}(\Theta_3)^\alpha (\Theta_3)_\alpha,
\end{equation}
\begin{equation}
    (X_3)_{\alpha\beta}
=
-( x_{13}^{-1})_{\alpha}{}^{\gamma}
( x_{12})_{\gamma\delta}
( x_{23}^{-1})^{\delta}{}_{\beta},\quad \quad \quad X_3^2 = -\frac{1}{2}(X_3)^{\alpha \beta} (X_3)_{\alpha \beta},
\end{equation}
as well as all the other possible index permutations.
One can show that the combinations
\begin{equation}
   \frac{\Theta_1^2}{\sqrt{X_1^2}}=\frac{\Theta_2^2}{\sqrt{X_2^2}}= \frac{\Theta_3^2}{\sqrt{X_3^2}}
\end{equation}
are all equal and superconformal invariants. Therefore, the most general form for the three-point function of superconformal primaries is given by
\begin{equation}\label{eq:3pt}
    \langle \Phi_i (z_1) \Phi_j (z_2) \Phi_k (z_3) \rangle = \frac{\lambda^E_{ijk}+\lambda^O_{ijk}\frac{\Theta_3^2}{\sqrt{X_3^2}} }{(y_{12}^2)^{\frac{\Delta_i+\Delta_j-\Delta_k}{2}} (y_{13}^2)^{\frac{\Delta_i+\Delta_k-\Delta_j}{2}} (y_{23}^2)^{\frac{\Delta_j+\Delta_k-\Delta_i}{2}}},
\end{equation}
where the $\Delta_l$'s are the conformal dimensions of the bottom scalar in each multiplet. The important point is that, for a given choice of superconformal primaries, \eqref{eq:3pt} contains only two independent structures, with opposite parity. They are parametrised by the two couplings $\lambda^E$ and $\lambda^O$, referring to the even and odd sector respectively. In a parity invariant theory, only one of the two can appear in three-point functions of superprimaries with definite parity. 

For three conformal primaries $\mathcal{O}_i,\mathcal{O}_j,\mathcal{O}_k$, we will use the standard notation $C_{ijk}$ for their OPE coefficient, defined through
\begin{equation}
    \langle \mathcal{O}_{i} (x_1)  \mathcal{O}_{j} (x_2) \mathcal{O}_{k}(x_3)\rangle=  \frac{C_{ijk}}{(x_{12}^2)^{\frac{\Delta_i+\Delta_j-\Delta_k}{2}} (x_{13}^2)^{\frac{\Delta_i+\Delta_k-\Delta_j}{2}} (x_{23}^2)^{\frac{\Delta_j+\Delta_k-\Delta_i}{2}}}.
\end{equation}Three-point functions of conformal primaries contained within the $\Phi$ multiplet can be obtained by projecting \eqref{eq:3pt} in different directions along superspace. As an example, one can extract the correlator of three super-primaries
\begin{equation}
\begin{split}
    \langle \mathcal{O}_{B,i} (x_1)&  \mathcal{O}_{B,j} (x_2) \mathcal{O}_{B,k}(x_3)\rangle=  \langle \Phi_i (z_1) \Phi_j (z_2) \Phi_k (z_3) \rangle \big \lvert_{\theta_1=\theta_2=\theta_3=0} =\\ & \frac{\lambda^E_{ijk}}{(x_{12}^2)^{\frac{\Delta_i+\Delta_j-\Delta_k}{2}} (x_{13}^2)^{\frac{\Delta_i+\Delta_k-\Delta_j}{2}} (x_{23}^2)^{\frac{\Delta_j+\Delta_k-\Delta_i}{2}}}.
    \end{split}
\end{equation}
Meanwhile, expanding \eqref{eq:3pt} in powers of $\theta^2$, and using \eqref{eq:exp}, one obtains
\begin{equation}
\begin{split}
    \langle \mathcal{O}_{T,i} (x_1) \mathcal{O}_{T,j}(x_2) \mathcal{O}_{B,k} &(x_3)\rangle=  \langle \Phi_i (z_1) \Phi_j (z_2) \Phi_k (z_3) \rangle \big \lvert_{\theta_1^2,\theta_2^2}= \\
    \frac{\left(\Delta_i+\Delta_j-\Delta_k\right)\left(\Delta_i+\Delta_j-\Delta_k-1\right)}{\sqrt{2 \Delta_i\left(2 \Delta_i-1\right)} \sqrt{2 \Delta_j\left(2 \Delta_j-1\right)}}\, & \frac{\lambda_{ijk}^E}{{(x_{12}^2)^{\frac{\Delta_i+\Delta_j-\Delta_k}{2}+1} (x_{13}^2)^{\frac{\Delta_i+\Delta_k-\Delta_j}{2}} (x_{23}^2)^{\frac{\Delta_j+\Delta_k-\Delta_i}{2}}}}.
    \end{split}
\end{equation}
Comparing the two equations, one infers 
\begin{equation}\label{eq:TTB}
C_{T_i T_j B_k}=\frac{\left(\Delta_i+\Delta_j-\Delta_k\right)\left(\Delta_i+\Delta_j-\Delta_k-1\right)}{\sqrt{2 \Delta_i\left(2 \Delta_i-1\right)} \sqrt{2 \Delta_j\left(2 \Delta_j-1\right)}}\, C_{B_i B_j B_k}.
\end{equation}
Similarly, one can show that
\begin{equation}\label{eq:TTT}
  C_{T_iT_jT_k}
  = \frac{
  (\Delta_i+\Delta_j+\Delta_k-3)
  (\Delta_i+\Delta_j+\Delta_k-2)
  }{
  \sqrt{2\Delta_j(2\Delta_j-1)}
  \sqrt{2\Delta_k(2\Delta_k-1)}
  }\,C_{T_iB_jB_k}.
\end{equation}
The relations \eqref{eq:TTB} and \eqref{eq:TTT}  also account for the proportionalities
between three-point functions observed in the explicit
holographic calculations of~\cite{Bobev:2025yxp}, providing a
superconformal explanation for relations that are less
transparent in the bulk description.

Moreover, if the theory is parity invariant, only one of the two structures can survive for a correlator involving super-primaries with definite parity. For a parity even correlator,
\begin{equation}
    C_{T_i B_j B_k}=C_{T_i T_j T_k}= 0,
\end{equation}
while for a parity odd one
\begin{equation}
      C_{B_i B_j B_k}= C_{T_i T_j B_k}=0.
\end{equation}

\bibliographystyle{JHEP}
\bibliography{holoconsist}

\end{document}